\documentclass[acmsmall]{acmart}
\AtBeginDocument{%
  }

\usepackage{tikz}
\usetikzlibrary{positioning,fit,arrows.meta,shadows,backgrounds}
\usepackage{booktabs}  %
\usepackage{tabularx}
\usepackage{array}     %
\usepackage{graphicx}  %
\usepackage{multirow}
\usepackage[most]{tcolorbox}
\usetikzlibrary{arrows.meta, positioning, shapes.multipart, decorations.pathreplacing,shapes, calc, backgrounds}
\usepackage{makecell}
\usepackage{tcolorbox}
\usepackage{booktabs}
\usepackage{array}

\newcolumntype{L}[1]{>{\raggedright\arraybackslash}p{#1}}
\newcolumntype{C}[1]{>{\centering\arraybackslash}p{#1}}

\tikzset{
    box/.style={draw, rounded corners=2pt, align=center, minimum width=3.8cm, minimum height=1.2cm},
    lightblue/.style={fill=cyan!10},
    lightorange/.style={fill=orange!10},
    lightpink/.style={fill=pink!10},
    lightpurple/.style={fill=purple!10},
    dottedarrow/.style={->, dotted, thick, draw=purple},
    arrow/.style={->, thick}
}

\usepackage{xcolor}
\definecolor{mBg}{RGB}{248,249,251}

\newcommand{\wei}[1]{\textcolor{black}{#1}}
\setcopyright{acmlicensed}
\copyrightyear{2018}
\acmYear{2018}
\acmDOI{XXXXXXX.XXXXXXX}
\acmConference[Conference acronym 'XX]{Make sure to enter the correct
  conference title from your rights confirmation email}{June 03--05,
  2018}{Woodstock, NY}
\acmISBN{978-1-4503-XXXX-X/2018/06}

\begin{document}

\title{Rethinking Testing for LLM Applications: Characteristics, Challenges, and a Lightweight Interaction Protocol}

\author{Wei Ma}
\email{weima@smu.edu.sg}
\affiliation{%
  \institution{Singapore Management University}
  \country{Singapore}
}

\author{Yixiao Yang}
\affiliation{%
  \institution{Capital Normal University}
  \city{Beijing}
  \country{China}
}

\author{Qiang hu}
\affiliation{%
  \institution{Tianjin University}
  \city{Tianjin}
  \country{China}
}

\author{Shi Ying, Zhi Jin and Bo Du}
\affiliation{%
  \institution{Wuhan University}
  \country{China}
}

\author{Zhenchang Xing}
\affiliation{%
  \institution{CSIRO's Data61 \& Australian National University}
  \country{Australia}
}

\author{Tianlin Li, Junjie Shi and Yang Liu}
\affiliation{%
  \institution{Nanyang Technological University}
  \country{Singapore}
}

\author{Lingxiao Jiang}
\affiliation{%
  \institution{Singapore Management University}
  \country{Singapore}
}

\renewcommand{\shortauthors}{W. Ma et al.}

\begin{abstract}
Applications of Large Language Models~(LLMs) have evolved from simple text generators into complex software systems that integrate retrieval augmentation, tool invocation, and multi-turn interactions. Their inherent non-determinism, dynamism, and context dependence pose fundamental challenges for quality assurance. This paper decomposes LLM applications into a three-layer architecture: \textbf{\textit{System Shell Layer}}, \textbf{\textit{Prompt Orchestration Layer}}, and \textbf{\textit{LLM Inference Core}}. \wei{We then assess the applicability of traditional software testing methods in each layer: directly applicable at the shell layer, requiring semantic reinterpretation at the orchestration layer, and necessitating paradigm shifts at the inference core. A comparative analysis of Testing AI methods from the software engineering community and safety analysis techniques from the AI community reveals structural disconnects in testing unit abstraction, evaluation metrics, and lifecycle management.} We identify four fundamental differences that underlie \wei{6} core challenges.  To address these, we propose four types of collaborative strategies (\emph{Retain}, \emph{Translate}, \emph{Integrate}, and \emph{Runtime}) and explore a closed-loop, trustworthy quality assurance framework that combines pre-deployment validation with runtime monitoring. Based on these strategies, we offer practical guidance and a protocol proposal to support the standardization and tooling of LLM application testing. \wei{We propose a protocol \textbf{\textit{Agent Interaction Communication Language}} (AICL) that is used to communicate between AI agents. AICL has the test-oriented features and is easily integrated in the current agent framework.}
\end{abstract}

\keywords{Large Language Models; Software Testing; AI System Testing; Runtime Testing; Testing Paradigm Evolution}

\maketitle

\definecolor{introboxcolor}{HTML}{303285}

\newcommand{\introbox}[1]{
  \begin{tcolorbox}[enhanced, left=3mm,right=3mm,
    colback=introboxcolor!10, colframe=introboxcolor!80, boxrule=0pt,
    borderline west={4pt}{0pt}{introboxcolor!90},
    ]
    #1
    \end{tcolorbox}
}

\section{Introduction}
\label{sec:introduction}
Applications of Large Language Models (LLMs) are reshaping the way software systems are constructed.
With the advent of technologies such as the Model Context Protocol (MCP)\footnote{\url{https://www.anthropic.com/news/model-context-protocol}}
, LLMs have evolved beyond text generation into application systems that support tasks such as retrieval-augmented question answering (Q\&A)\cite{wiratunga2024cbr} and code generation\cite{jiang2024survey}.
These systems are no longer simple language generators but complex intelligent software systems~\cite{shimtooldial,wang-etal-2025-rethinking-stateful,mcnamara2025demystifying} that integrate retrieval-augmented generation (RAG), tool invocation, multi-turn dialogues, and user interactions.

As LLM applications evolve from single-turn Q\&A to interaction modes involving multi-agent orchestration~\cite{mialonaugmented}, chain-of-thought (CoT) reasoning~\cite{wei2022chain}, and external API calls, their runtime behavior exhibits greater variability and is more prone to anomalies.
Compared to traditional software systems, LLM applications exhibit distinct characteristics~\cite{atf2025challenge}: non-determinism, dynamic behavior, and context dependence.
Their outputs arise from the interplay of prompt formulation, contextual state, and model version.
Small perturbations in prompt or context can lead to noticeable changes in outputs, a phenomenon loosely analogous to sensitivity to initial conditions in chaos theory~\cite{11030017}, while LLM applications are not chaotic systems in the strict scientific sense.

Quality assurance for LLM applications has become both pressing and multifaceted.
In traditional software engineering, testing is a cornerstone of quality assurance.
Faced with this new paradigm, the question is not whether testing should be abandoned, but how it should be adapted and retained as a primary quality assurance mechanism.
Before discussing methodologies, we first analyze the \emph{fundamental differences} between LLM applications and conventional software.
These differences underlie and often give rise to the distinctive testing challenges we face.

\subsection{Differences Between LLM Applications and Traditional Software}
\label{sec:fundamental}
Traditional software systems are grounded in deterministic algorithmic logic and, under identical execution environments and input, typically produce predictable and repeatable outputs~\cite{promptedmind2025deterministic, alphanome2025probabilistic}. 
Although non-deterministic behaviors may occasionally arise, for example in cases involving concurrency or reliance on external services, they are usually controllable, analyzable, or mitigated through engineering mechanisms such as synchronization protocols or retry strategies.
By contrast, LLM applications reflect shifts: \textit{from deterministic computation to probabilistic generation}~\cite{acm2024nondeterminism, barton2023challenge}, \textit{from purely logical execution to brain-inspired modeling paradigms}\cite{shelton2023computing, nature2019deterministic, oxford2024advancing}, and \textit{from static rule-based systems to dynamically learned behaviors}~\cite{arxiv2024braininspired, mdpi2021incremental}.
These shifts manifest in several key aspects:
\begin{itemize}
  \item \textit{Open-ended output space}: Traditional software typically produces outputs from a predefined, finite set, whereas LLM applications operate over an open natural language space, where the same input may yield multiple semantically valid outputs. This weakens traditional exact-match verification and motivates semantic evaluation methods, such as LLM-as-judge approaches.
  \item \textit{Context-dependent behavior}: Responses from LLM applications depend not only on the current input but also on dialogue history, system state, and external knowledge. This context dependence exceeds the modeling capacity of traditional testing methods and calls for a shift from stateless function verification to stateful analysis of behavioral sequences.
  \item \textit{Emergent and difficult-to-localize capabilities}: Certain LLM abilities (e.g., multi-step reasoning, creative writing) emerge during large-scale training and cannot be solely validated by code logic. Their boundaries and reliability are primarily assessed through \emph{empirical evaluation} rather than \emph{formal verification}.
  \item \textit{System-level composition and complexity}: LLM applications integrate multiple subsystems such as retrieval augmentation, tool invocation, and multi-turn interaction, yielding architectures more complex than traditional software. Interactions among these components introduce new failure modes and additional testing requirements.
\end{itemize}

These fundamental differences help explain when traditional approaches may fail and indicate where innovations in LLM testing are most needed.
At its core, testing is an \textbf{observational activity}: \textit{a systematic process for examining system behavior under varied conditions}.
Regardless of methodology, any quality assurance approach for LLM applications should involve such systematic observation.
The necessity of testing rests on a simple premise: any deployed system needs to be understood and validated.
This requirement applies to both deterministic and probabilistic systems.
Indeed, the inherent uncertainty of LLM applications makes systematic behavioral observation \textit{more important}, not less.

The key question, therefore, is not whether testing is needed, but how methodologies should be adapted or newly designed to align with the characteristics of LLM applications.
Completely discarding traditional methods is neither realistic nor necessary.
The conventional code and integration logic of an LLM application can still be tested with established techniques, whereas the model inference core requires complementary, AI-specific evaluation approaches.
Thus, traditional and novel approaches should coexist, each applied where they are most effective.
Our work examines how traditional software testing can be integrated with AI-specific evaluation to address the unique challenges of assuring LLM applications.

\subsection{Architectural Model of LLM Applications}
\label{sec:architectural}
The three-layer decomposition we propose in this work is not ad hoc, but a structural pattern consistently observed in both academic research and industry practice.
Frameworks such as the 12-Factor Agents blueprint~\cite{humanlayer2024twelvefactor}, LangGraph~\cite{langgraph2024overview}, and Microsoft AutoGen~\cite{autogen2023microsoft} exhibit a similar separation among (i) integration and runtime components, (ii) orchestration logic, and (iii) model inference.
Comparable distinctions are also documented in surveys of LLMOps and AgentOps~\cite{yehudai2025survey, mcnamara2025demystifying}. 
This separation reflects clear boundaries in determinism, control, and observability across layers, which correspond to fundamentally different testing focuses.
This convergence between independent engineering practices and academic analyses provides a conceptually sound and practically grounded foundation for the subsequent layer-specific discussion of testing.
To show system-level behavior in LLM applications and in line with the principle of \textit{separation of concerns}, we propose three functional layers. We then use this structure to situate the scope of testing methods and organize their coordination.

\begin{enumerate}
    \item \textit{System Shell Layer}: The peripheral support components surrounding the LLM core, with an emphasis on \emph{runtime and integration}. Typical elements include API interfaces, preprocessing and postprocessing logic, tool invocation modules, human-in-the-loop interfaces, event triggers, and reliability mechanisms such as retries and timeouts.

    \item \textit{Prompt Orchestration Layer}: Dynamically composes final prompts from user inputs, system instructions, external knowledge, and contextual information. It manages stateful control flow, context compression, and multi-agent handoffs, as exemplified by orchestration frameworks such as LangGraph~\cite{langgraph2024overview}, which support persistent execution state and graph-structured workflows.

    \item \textit{LLM Inference Core}: Includes the model parameters, inference services, decoding strategies, and safety mechanisms. As the probabilistic generation engine, its outputs depend on prompts, context, and configuration, and it is typically treated as a black box~\cite{mukherjee2024methodology}.
\end{enumerate}

This structure clarifies the scope and applicability of testing techniques at each layer and underscores the need for cross-layer analysis. This architectural view is independent of any particular testing agenda; subsequent sections leverage it to position and coordinate testing methods.
From the \emph{System Shell} to the \emph{LLM Inference Core}, \textit{behavioral predictability decreases progressively}, whereas \textit{testing complexity and cost increase correspondingly}~\cite{yehudai2025survey}.
For example, unit and integration testing remain effective for shell components such as API interfaces and data preprocessing. The orchestration layer benefits from combining traditional logic validation with semantic quality evaluation, whereas the probabilistic nature of the inference core calls for distinct, probabilistic evaluation paradigms.

Equally important is \emph{inter-layer interaction testing}.
In practice, many faults arise not from individual layers but from data flow and state propagation across layers.
For instance, a formatting error in the shell layer may cause the orchestration layer to generate an incorrect prompt. This can then trigger abnormal outputs from the inference core and ultimately degrade system-level behavior~\cite{mukherjee2024methodology}.
Traditional testing methods may overlook such cross-layer error propagation, whereas AI evaluation methods often provide limited modeling of system-wide interactions.

\subsection{Research Objectives and Analytical Framework}
\label{sec:objectives}
This paper proceeds along the following objectives:
\begin{enumerate}
\item Applicability of traditional testing (Section~\ref{sec:ttm}). Using the three-layer structure, we analyze where classical testing methods remain applicable (e.g., interface validation and shell-layer logic) and where they face limitations.

\item Survey of AI-oriented evaluations (Section~\ref{sec:att}). We review AI-oriented evaluation methods in software engineering and assess their alignment with the demands of LLM applications. We also consider complementary AI-safety techniques (privacy attacks, model extraction, data-poisoning detection, and bias assessment) in the context of quality assurance.
  
\item Challenge identification and classification (Section~\ref{sec:core-challenges}). Starting from model non-determinism and system composition, we identify the primary challenges in testing LLM applications and organize six core issues into a multi-dimensional view.

\item Collaborative framework design (Section~\ref{sec:collab}). We outline how multiple methods can be positioned and integrated into a comprehensive assurance workflow. We also propose a lightweight structured protocol for testable LLM agent systems to enable replay and comparison in evaluation.
\end{enumerate}

\section{Applicability and Evolution of Traditional Testing Paradigms}
\label{sec:ttm}
Faced with the differences between LLM applications and traditional software, a natural question is whether traditional software testing methods remain effective. While the non-deterministic nature and semantic complexity of LLMs introduce new challenges, LLM applications still consist largely of conventional code. Unit, integration, and performance testing, together with other conventional techniques, remain foundational to software quality assurance. More precisely, traditional testing approaches are not obsolete; they need to be repositioned within the new technical landscape. This section uses the three-layer architectural model to examine the applicability boundaries and evolutionary paths of traditional testing paradigms.

Based on the three-layer architectural model introduced in Section~\ref{sec:architectural}, we classify the roles of traditional testing methods in LLM applications into three categories:

\begin{itemize}
  \item \textbf{Retained applications}: methods directly applicable to deterministic components.
  \item \textbf{Translatable adaptations}: methods that require redefining test units and objectives.
  \item \textbf{Integrative innovations}: methods that require semantic awareness and extensions for behavioral modeling.
\end{itemize}

This classification clarifies appropriate usage scenarios and provides a structured roadmap for evolving traditional testing approaches in the LLM era.

\subsection{Retainable Role}
Within the three-layer architectural model of LLM applications, the \textit{System Shell Layer} follows the classical input-process-output paradigm and features well-defined functional specifications and state-transition logic.
Existing techniques ensure logical correctness and operational stability in this layer.
Consequently, traditional testing methods are highly applicable at this level. Specifically:

\begin{itemize}
  \item Unit testing. Verifies the behavior of non-model modules and ensures correct execution of function-level routines such as format parsing and field extraction.

  \item Interface testing. Validates request/response schemas, retry mechanisms, and exception-handling logic. This is especially critical when integrating third-party LLM APIs (e.g., OpenAI, Claude).

  \item Integration testing. Checks cross-component interactions and ensures end-to-end functionality from user input through the retrieval component to LLM inference.

  \item Performance and stress testing. Evaluates service-level metrics such as response latency, memory footprint, and GPU utilization under high concurrency, multi-turn interactions, and large-input scenarios.
\end{itemize}
These traditional methods provide a foundation for engineering LLM applications. They directly support reliability and maintainability and help establish a stable operational baseline.

\subsection{Translatable Role}
Although the core concepts of traditional testing are not directly applicable to LLM applications in their original form, the underlying principles of adequate coverage, behavioral consistency, and regression prevention remain valuable.
In LLM applications, these principles should be reinterpreted using semantic behavior modeling.
For example, assertion-based testing of the form ``\verb|expected == actual|'' faces fundamental challenges in LLM applications because outputs are natural-language sequences characterized by multiplicity and subjectivity.
Boolean assertions should be augmented or replaced by mechanisms such as LLM-based scoring, semantic-similarity measures, or multi-candidate comparison:

\begin{itemize}
  \item LLM-as-judge scoring: this approach is easy to implement and widely used, but it requires monitoring scorer consistency and potential bias.
  \item Semantic-similarity evaluation: compute cosine similarity using models such as BERT or Sentence-Transformers. A key challenge is setting appropriate similarity thresholds.
  \item Multi-candidate voting: generate multiple candidate outputs and use an expert-agent system to select the best answer.
\end{itemize}

Translating coverage testing entails shifting from logical paths to semantic paths.
Traditional notions such as path and branch coverage have limited validity given the black-box nature of LLM inference.
Prompts define linguistic rather than logical paths.
Accordingly, new coverage metrics should be defined:

\begin{itemize}
  \item Prompt-embedding coverage. Map prompts to a high-dimensional embedding space and use clustering to ensure that test cases cover diverse semantic regions.
  \item Semantic-category coverage. Define a compositional space of prompt templates to ensure adequate coverage of task types, linguistic styles, and complexity levels.
  \item Output-behavior diversity. Use entropy-based indicators to assess the richness of output distributions and reduce mode collapse.
\end{itemize}

Regression testing should also be reinterpreted in terms of behavioral consistency.
Traditional output-equivalence checks are often unreliable, because identical inputs can yield divergent outputs due to context, caching, or versioning.
Therefore, new strategies for evaluating behavioral consistency are needed.
For example, in code-generation stability testing, using 100 semantically equivalent function descriptions (e.g., ``sort a list'' versus ``arrange list elements in order''), the model is expected to achieve at least 95\% functional correctness with acceptable stylistic variation.

This translational strategy embodies the core idea of semantic behavior modeling, which shifts the test unit from static code paths to observable linguistic behavior patterns, including output distributions, semantic categories, and context-driven variation.

\subsection{Integrative Role}
Traditional testing methods primarily focus on functional correctness, whereas the testing objectives for LLM applications also include content plausibility, contextual consistency, factual accuracy, bias, and safety.
Achieving these goals requires incorporating methods from natural language processing, AI safety, and human-computer interaction testing to construct a cross-disciplinary, integrated framework.

For output robustness, traditional approaches often overlook linguistic variation and ambiguity.
In contrast, LLM testing should employ semantics-preserving prompt rewrites to evaluate output variance and stability across paraphrases.
Factual and knowledge verification is not a primary focus of traditional testing.
LLM testing should incorporate retrieval-augmented checkers and time-sensitive datasets to assess hallucinations, staleness, and factual drift.
Testing contextual-state consistency involves constructing context-state trees and analyzing prompt-behavior consistency and transition patterns across alternative context paths, which may exceed the modeling capacity of traditional user-scenario testing.
Moreover, for domain-specific LLM applications, constructing a knowledge graph of real-world facts may serve as a way to approximate semantic test coverage.
However, building such a knowledge base is nontrivial.

This category of testing gives rise to a new research direction that aims to build a complete testing loop spanning semantic generation, behavioral observation, and contextual memory. This integrative role indicates that LLM application testing is not merely an extension of existing techniques but also entails a transformation in testing philosophy and objectives.

Traditional software testing remains valuable and provides a foundation for quality assurance in LLM applications. However, as discussed earlier, traditional approaches face notable limitations in both the translatable and integrative roles, particularly in semantic evaluation, behavior modeling, and contextual-state management. These challenges motivate ongoing development of testing methods specifically tailored to AI systems.

It is therefore worth examining whether emerging AI-oriented testing methodologies can effectively address the gaps left by traditional testing. In the following section, we examine the current landscape and limitations of AI testing research and assess how closely these efforts align with the needs of LLM applications.

\section{Current Landscape and Capability Boundaries of AI Testing Research}
\label{sec:att}

Both the software engineering and AI research communities are actively exploring new testing theories and methodologies. However, an open question is whether these efforts address the quality-assurance needs of LLM applications and whether structural disconnects exist between the approaches taken by the two communities.

Current research on AI testing has evolved into two largely independent directions: one is \textit{Testing AI} from a software engineering perspective, and the other is \textit{security analysis techniques} rooted in AI safety. The former emphasizes functional correctness and engineering quality, whereas the latter targets model security risks and ethical considerations.

\paragraph{\textbf{Category 1:} Testing AI (software engineering perspective)}
Researchers in the software engineering (SE) community consider AI models as software components and focus on functional correctness, reliability, and engineering quality~\cite{zhang2020machine}. The core objectives are to verify that AI systems operate as expected, to detect functional flaws, and to evaluate system performance. This line of work builds on and extends traditional theories, methods, and tools in software testing.

\paragraph{\textbf{Category 2:} AI security analysis (AI research perspective)}
Researchers in the AI community focus on safety, privacy, fairness, and ethical risks associated with AI models. Their goals are to identify vulnerabilities, privacy leaks, algorithmic bias, and adversarial weaknesses~\cite{hu2021artificial}. This line of work builds on subfields such as machine learning security, privacy-preserving learning, and algorithmic fairness.

While both categories have made significant progress in their respective domains, each faces notable limitations when addressing the full complexity of LLM applications. More critically, differences in focus and methodology between the two communities have led to a structural disconnect in addressing quality assurance for LLM applications.

\subsection{Testing AI from SE}
AI testing research in the software engineering community has primarily focused on extending established theories and methodologies from traditional software testing to AI systems. The central idea is to treat AI models as software components with distinctive properties. Under this paradigm, mainstream \textit{Testing AI} approaches generally follow three technical pathways:

\paragraph{(1) Input perturbation and robustness testing}
This category draws on mutation testing by introducing character-level, word-level, or syntactic perturbations to inputs to generate adversarial or variant test cases. The objective is to verify whether outputs remain semantically consistent. This assumes that invariant input semantics correspond to invariant output semantics, using input stability as a proxy for output robustness. These methods are mainly applicable to static-input tasks such as text classification, sentiment analysis, and single-turn Q\&A. However, they are limited for multi-turn context, state evolution, external tool calls, and prompt chaining, and they offer little support for modeling complex prompt assembly and contextual memory.

DeepXplore~\cite{pei2017deepxplore} pioneered white-box testing of neural networks, using differential testing to identify inconsistent DNN behaviors. Tools such as TensorFuzz~\cite{odena2019tensorfuzz} and DeepHunter~\cite{10.1145/3293882.3330579} apply coverage-guided fuzzing. Frameworks such as CheckList~\cite{ribeiro2020beyond} offer systematic behavioral testing for NLP models. Tools including TextFooler~\cite{jin2020bert}, DeepWordBug~\cite{gao2018black}, and TextBugger~\cite{li2019textbugger} focus on adversarial text generation. The DeepMutation series~\cite{ma2018deepmutation} extends mutation testing to deep learning. Recent open-source frameworks such as DeepEval\footnote{\url{https://github.com/confident-ai/deepeval}} attempt to integrate traditional testing techniques with AI-safety analyses, offering unified interfaces and metrics. However, they still face limitations in modeling system-level behavior and in reducing reliance on external evaluators.

\paragraph{(2) Metamorphic testing}
Metamorphic testing\footnote{\url{https://en.wikipedia.org/wiki/Metamorphic\_testing}} defines relations between inputs and outputs, known as metamorphic relations (MRs), to evaluate behavioral consistency when ground-truth answers are unavailable. Its core value is enabling large-scale automated test generation and supporting multidimensional evaluation, including diversity, robustness, and fairness.

In practice, metamorphic testing has shown promise across several domains. For example, DeepRoad~\cite{zhang2018deeproad} applies GAN-based image transformations for autonomous-driving system testing, and in NLP testing researchers use language-preserving transformations such as synonym substitution and syntactic rewriting to verify consistency. Tools such as METAL~\cite{10638599} and MORTAR~\cite{guo2024mortar}, together with prompt-level semantic metamorphic testing, represent subsequent developments of this approach. Nevertheless, metamorphic-testing methods face clear limitations. The design of metamorphic relations often relies on manual heuristics, increasing complexity and limiting scalability to the intricate prompt paths, tool-call sequences, and memory-dependent behaviors present in LLM applications.

\paragraph{(3) Neural probing and coverage modeling}
Neural probing and coverage modeling seek to estimate test-sample coverage by analyzing neuron activations, attention patterns, or distributional clustering. The goal is to guide test generation and behavioral detection. The core idea is to transfer coverage concepts from traditional software testing to deep neural models to estimate test adequacy when ground-truth labels are unavailable.

This line of work has produced several technical branches:  
\textit{Structural coverage} focuses on neuron and branch coverage as well as boundary metrics~\cite{usman2023overview}.  
\textit{Behavioral coverage}, such as Surprise Adequacy~\cite{10.1145/3546947}, measures adequacy based on behavioral distributions.  
\textit{Symbolic verification} methods such as Reluplex~\cite{katz2017reluplex} employ SMT solvers for formal verification of neural networks.  
\textit{Concolic testing}~\cite{sun2018concolic} extends symbolic execution techniques to DNNs.
However, these methods face practical obstacles. The internal states of deep models are high-dimensional and lack standardized representational frameworks, and, most critically, correlations between coverage and actual functional defects are often weak. This limits their applicability in real-world testing.

\subsection{AI Security Analysis Methods}
Security analysis methods in the AI research community arise from investigations into the intrinsic characteristics of machine learning models. These methods primarily focus on internal security risks, ethical considerations, and broader social implications~\cite{hu2021artificial,rahman2023security}. Originally, the goal was to assess properties (\textit{security, privacy, fairness, and robustness}) that differ fundamentally from the \textit{functional-correctness verification} emphasized in the software engineering community~\cite{das2025security,yao2024survey}.

From the perspective of LLM application testing, these methods constitute a core component of the quality-assurance landscape, especially for addressing semantic safety, behavioral consistency, and ethical risk that traditional functional testing may not cover. Table~\ref{tab:quality} summarizes the categories and value of AI security analysis methods for testing LLM applications.

\begin{table}[]
\centering
\caption{Quality assurance value of AI security analysis methods}
\small
\label{tab:quality}
\begin{tabular}{|p{3cm}|p{5cm}|p{5cm}|}
\hline
Analysis dimension & Representative methods & Core role in LLM application testing \\
\hline
Training data privacy & Membership inference~\cite{hu2022membership}, model inversion~\cite{fredrikson2015model} & Privacy risk assessment and compliance-oriented testing \\
\hline
Model intellectual-property protection & Black-box model extraction~\cite{yue2021black}, distillation~\cite{xu2024survey}, output watermarking~\cite{pan2024markllm} &IP risk assessment and ownership attribution/tracing \\
\hline
Data poisoning & Poisoning-sample insertion~\cite{zhu2023boosting}, backdoor injection~\cite{yan2024backdooring} & Detection of abnormal behavior during fine-tuning \\
\hline
Deployment and supply-chain security & LoRA weight sideloading~\cite{yao2024risks}, KV-cache leakage~\cite{wu2025know}, third-party component injection~\cite{zhao2025a} & Multi-tenant inference isolation and supply-chain integrity \\
\hline
Bias and fairness & Multi-attribute bias QA~\cite{parrish2022bbq}, holistic bias prompts~\cite{smith2022holistic} & Fairness assurance across demographic groups \\
\hline
Adversarial prompt attacks & Jailbreak-prompt generation~\cite{chao2023jailbreak}, indirect prompt injection~\cite{abdelnabi2023indirect} & Robustness testing of safety alignment under malicious or adversarial prompts \\
\hline
Output explainability probes & Attention-weight tracing~\cite{artzy2024attention}, knowledge-neuron mapping~\cite{dai2022knowledge} & Behavior attribution and debugging in complex scenarios \\
\hline
Watermarking and output tracing & Invisible-text watermarking (SynthID)~\cite{dathathri2024synthid}, robust watermark detection (Tr-GoF)~\cite{li2024trgof} & Ownership verification and auditing of generated outputs \\
\hline
\end{tabular}
\end{table}

\paragraph{Layer-Specific Applicability of AI Security Analysis Methods}
Although AI security analysis methods are a core component of LLM application testing, their applicability is mostly confined to specific architectural layers. Traditional software testing primarily targets the system shell layer, focusing on functional correctness and performance stability. In contrast, AI security analysis methods are designed to address the unique challenges posed by the prompt orchestration layer and the LLM inference core, namely, semantic safety, behavioral consistency, and ethical risks, that are difficult to capture through traditional testing.

Current research suggests that AI security analysis methods are largely focused on the third layer (the LLM inference core) emphasizing internal model property evaluation. In contrast, the second layer, which involves complex prompt orchestration and context-state management, receives comparatively little attention. This imbalance in research emphasis leads to adaptation challenges when applying these methods in practical system-level scenarios.

\subsection{Structural Disconnection Between the Two Methodological Paradigms}
The divergence in methodologies, evaluation metrics, and technical pathways between the two communities has led to a structural disconnection in addressing the complex requirements of LLM applications. This disconnection reflects not only limitations of individual methods but, more importantly, the lack of unified theoretical frameworks and collaborative mechanisms.

\paragraph{\textbf{Structural limitations of Testing AI approaches}}
The software engineering community is accustomed to testing deterministic systems, and its testing theories rest on assumptions of reproducibility, controllability, and predictability. However, the non-deterministic characteristics of LLM applications are in tension with these assumptions. We identify four systemic mismatches between Testing AI approaches and the needs of LLM applications:

\begin{enumerate}
  \item 
  Mismatch in test unit abstraction.
  AI testing typically treats the ``input sample'' as the test unit, while the behavior of an LLM application system is a composite triggering structure composed of ``prompt chaining + state memory + multi-agent coordination''.
  In real systems, agent tasks often involve sequences such as ``goal decomposition → tool planning → execution feedback → state update → behavioral recursion'', which go far beyond the expressive scope of current testing paradigms.
  \item  Incompatibility between behavioral modeling assumptions and system execution paths.
  Existing tests are mostly based on a short-path model assumption of ``input → output'', while actual systems follow a heterogeneous pipeline consisting of ``prompt + retrieval + tool invocation + multi-turn state''.
  In practice, prompt perturbation tests are often neutralized by concatenation mechanisms, filtering rules, system prompts, or session memory, limiting their effectiveness.
  Moreover, behaviors of key components such as prompt orchestrator, and tool states are often unobservable.
  Current testing methods cannot trace the generation path, resulting in unexplainable test outcomes, undiagnosable errors, and unreproducible regressions.

  \item Lack of multi-dimensional semantic evaluation metrics.
  Commonly used metrics such as BLEU, coverage, and clustering distance offer limited discriminative power.
  These methods suffer from major deficiencies and they lack a behavioral classification system and dimension-specific metrics, making feedback-driven optimization difficult.

  \item Missing lifecycle testing pipeline.
  Most research focuses on pre-deployment testing, while real-world systems undergo continuous evolution, such as prompt strategy updates, LoRA fine-tuning iterations, and RAG index replacements;
  There is a lack of CI/CD-integrated mechanisms for version comparison testing, behavioral difference detection, and semantic regression verification;
  Test sets cannot incrementally grow from real-world anomalies and quickly become obsolete.
\end{enumerate}

\paragraph{\textbf{Adaptation Difficulties of AI Security Analysis Methods}}
Existing AI security analysis methods face notable adaptation challenges when applied to real-world LLM systems, primarily due to differences between research assumptions and production environments. Many of these methods rest on an idealized premise that treats the model as the sole object of study, emphasizing internal properties and theoretical bounds while overlooking interaction effects and state dependencies in complex system environments. In practice, prompt data in LLM applications typically flows through multi-stage transformation pipelines, which often undermines the effectiveness of attacks designed for static, single-turn inputs. Moreover, features such as memory mechanisms, session-state caching, and multi-agent collaboration cause responses to depend not only on the current input but also on historical outputs, forming intricate indirect causal chains. This dynamic state evolution can expose and amplify security risks over multi-turn interactions. However, traditional static evaluation, limited to a single point in time, may fail to capture this accumulation of risk. As a result, evaluation outcomes derived from isolated experiments may not generalize well to production environments characterized by multi-component coordination, shared state, and concurrent access.

Looking ahead, the capability boundaries and disconnects outlined above define the problem space we address next: Section~4 organizes testing pain points into system-level challenge categories grounded in the architectural view introduced in Sections~\ref{sec:architectural}.

\section{Core Challenges in Testing LLM Applications}
\label{sec:core-challenges}

Building on the fundamental differences in Section~1.1, the three-layer architectural view in Section~1.2, and the capability boundaries identified in Section~3, we consolidate the problem space into system-level challenge categories for testing LLM applications. Each category is defined at a granularity beyond a single prompt–response pair, aligned with the architecture, and motivated by the issues discussed above.

\subsection{Challenges}
\subsubsection{Semantic Evaluation and Behavioral Consistency}
Outputs of LLM applications lie in an open-ended natural-language generation space, where a single input may admit multiple semantically correct answers. Traditional exact-match verification methods are therefore insufficient on their own in this context. Moreover, conventional coverage metrics (e.g., statement or branch coverage) do not capture the breadth of the semantic behavior space, leaving few reliable quantitative measures of behavioral adequacy.  

Compounding the difficulty, LLMs exhibit marked sensitivity to superficial linguistic variation, such that minor, semantically equivalent rephrasings can trigger substantial fluctuations in output quality. As a result, testing efforts struggle to define a unified correctness standard, accurately measure coverage, and predict system behavior in production, which can affect user experience and system stability. Addressing these challenges calls for a shift from static path coverage to semantic behavior modeling, combined with techniques such as LLM-as-judge and semantic-similarity measures to evaluate output consistency and stability.

\subsubsection{Test Data and Robustness in an Open Input Space}
The input space of LLM applications is open and semantically unbounded. Real-world user inputs often contain typos, dialectal expressions, emotional language, or implicit assumptions, making full test coverage impractical. Furthermore, LLM boundary conditions are not simple numerical or format constraints but semantically defined extremes. Such atypical inputs can cause performance degradation, logical incoherence, or hallucinations.  

Constructing test sets that both represent real-world distributions and cover abnormal scenarios is costly and prone to rapid obsolescence. The combinatorial effects of different anomaly types can grow rapidly, making traditional boundary-value analysis insufficient. Therefore, robustness assessment calls for distribution-aware data generation and combinatorial testing of anomaly patterns to maintain ecological validity in evaluation.

\subsubsection{Dynamic State and Observability}
Many LLM systems are stateful. Historical context, key-value (KV) caches, and conversational memory can propagate across turns, leading to behavioral carryover and violations of user isolation. At the same time, LLM outputs are influenced by multiple layered components, including retrieval modules, prompt templates, and reasoning engines. However, many systems lack a unified tracing and observability framework, which makes failure reproduction and root-cause diagnosis challenging.  

Such state dependence and limited visibility undermine the basis for debugging, regression testing, and performance optimization. Mitigation strategies include introducing end-to-end observability frameworks with unified trace IDs and reconstructable context states, multi-turn conversation-replay mechanisms, and cross-component data-flow monitoring tools.

\subsubsection{Capability Evolution and Regression Risks}
After training, a model’s internal knowledge is effectively fixed and can become outdated over time. Task-specific fine-tuning can enhance targeted capabilities while potentially degrading overall performance. Without a standardized multi-task regression testing framework, capability degradation or regression is often detected only weeks or months after deployment.  

These risks affect not only time-sensitive applications (e.g., news, finance, and legal domains) but also lead to unpredictable behavioral differences in multi-version deployments. Addressing these risks calls for establishing multidimensional capability benchmarks, coupled with continuous regression testing and version-differential analysis, to enable early warning of knowledge obsolescence and capability drift.

\subsubsection{Security and Ethical Compliance Challenges}
LLMs can be sensitive to adversarial prompts (jailbreaks) and covert malicious instructions. Attackers may bypass safety mechanisms through role-playing, indirect prompt injection, or other prompt-manipulation techniques. Divergent cultural and regulatory definitions of “harmful” complicate cross-jurisdictional compliance.  

Latent risks include harmful content embedded in figurative language, culturally sensitive topics, and passively triggered malicious instructions. Legal accountability is often ambiguous, which calls for multi-layer safety defenses at the model, application, and user levels, together with dynamic safety evaluation and localized compliance strategies.

\subsubsection{Multimodal and System-Level Integration Challenges}
LLM applications increasingly integrate capabilities such as image understanding, speech recognition, video analysis, and multi-agent collaboration, yielding complex multimodal systems. Errors in one modality can be amplified or masked as they propagate through the pipeline, making them difficult to detect in single-module tests.  

Meanwhile, long-context processing, repeated tool invocations, and multi-agent coordination can introduce nonlinear increases in inference latency and resource consumption. Standardized methods for cross-modal consistency verification are lacking, and overall system performance is jointly influenced by semantic complexity, interaction depth, and call-chain length, complicating capacity planning and cost control. Addressing these issues calls for cross-modal consistency-checking frameworks and semantic-aware performance-testing methods, together with early-stage design mechanisms to balance performance, accuracy, and cost.

\subsection{The Inevitable Shift in Testing Paradigms}
\label{sec:shift}
This analysis indicates that LLM testing challenges are not merely technical or procedural; they signal a shift in testing philosophy:

\begin{itemize}
    \item \textit{From deterministic validation to probabilistic evaluation}: prioritize semantic equivalence, calibrated confidence intervals, and robustness analysis over exact-match checks.
    \item \textit{From static coverage to dynamic exploration}: adopt risk-driven, distribution-aware, and adversarial data generation in place of fixed test sets.
    \item \textit{From one-time validation to continuous assurance}: employ continuous monitoring, drift detection, and runtime regression to address model evolution and state-dependent behavior.
    \item \textit{From single method to collaborative ecosystem}: integrate semantic evaluation, robustness probing, multimodal consistency verification, and performance and cost optimization into a coordinated testing framework.
\end{itemize}

These shifts are not incremental improvements; they redefine the aims and instruments of testing and serve as guiding principles for assessing the applicability and evolution of methods in the sections that follow.

\section{Collaborative Testing Framework Design}
\label{sec:collab}
Based on the four paradigm shifts, we first outline a collaborative testing framework that integrates the strengths of diverse testing methodologies. We then introduce AICL, a lightweight structured protocol that operationalizes this framework by ensuring semantic precision, observability and provenance, deterministic replay, context isolation, and built-in evaluation hooks for fair comparison.

\subsection{Collaboration Strategies}
Given the six challenge categories consolidated above, no single testing method is sufficient on its own.
Traditional testing remains effective for deterministic components but lacks semantic evaluation.
AI analysis methods introduce innovations in behavioral modeling but remain underdeveloped for system-level engineering~\cite{yuan2023no}.
AI security analysis excels at risk identification yet is often disconnected from real-world operational contexts~\cite{liu2023prompt, wei2023jailbroken}.
Each approach exhibits distinct strengths and blind spots.

The core of a collaborative testing framework is to translate the four paradigm shifts into actionable testing strategies (Section~\ref{sec:shift}). This requires a unified quality-assurance system that coherently integrates the strengths of different methods while mitigating their limitations. Building on the six-category analysis, we illustrate the method applicability boundaries and design effective collaboration strategies.

\begin{table}[h]
\centering
\caption{
Applicability and Collaboration Strategies of Testing Methods Across Challenge Categories}
\label{tab:applicability}
\small
\scalebox{0.8}{
\begin{tabular}{|p{3.5cm}|p{3cm}|p{3cm}|p{4cm}|}
\hline
\textbf{Challenge Category} & \textbf{Traditional Testing} & \textbf{AI-oriented Evaluation} & \textbf{Collaborative Resolution Strategy} \\
\hline
Semantic Evaluation and Behavioral Consistency & Exact-match assertions invalid; code coverage inapplicable & Partially applicable (semantic scoring, embedding-based metrics) & Translational: semantic scoring + multi-candidate validation + prompt diversity \\
\hline
Test Data and Robustness in an Open Input Space & Partially applicable (boundary values, equivalence classes) & Partially applicable (adversarial input generation) & Hybrid: distribution-aware data synthesis + anomaly-pattern combination testing \\
\hline
Dynamic State and Observability & Limited state modeling; hard to control environment & Lacks unified state tracking, partial contamination detection & Runtime: end-to-end observability + state replay + semantic contamination detection \\
\hline
Capability Evolution and Regression Risks & Regression testing frameworks available & Limited capability benchmarking & Translational: multi-task capability regression + version-differential analysis \\
\hline
Security and Ethical Compliance & Blind spots in semantic risks; useful for infrastructure security & Specialized risk detection (prompt injection, harmful content) & Preservational: primarily AI security analysis, supplemented by system-level security testing \\
\hline
Multimodal and System-Level Integration & Interface testing applicable but lacks cross-modal scope & Limited cross-modal consistency modeling & Hybrid: end-to-end multimodal pipeline validation + semantic consistency checks \\
\hline
\end{tabular}}
\end{table}

From the mappings in Table~\ref{tab:applicability}, it follows that different challenges call for different collaborative strategies.
Based on similarities in resolution approaches, the six challenge categories can be addressed via four collaborative modes.

\paragraph{\textbf{1. Preservational application: extending the strengths of dominant methods}}
In some categories, one testing approach is primary and the other plays a complementary role.
For example, \textit{system-level performance} concerns in the \textit{multimodal and system-level integration} category can often be addressed with mature, traditional techniques for performance, stress, and load testing~\cite{iso25010_2023}.
Although LLM inference patterns differ from traditional software execution, bottlenecks, memory leaks, and concurrency issues can often be diagnosed with established tools.

By contrast, \textit{security and ethical compliance} relies primarily on AI security analysis. LLM-specific threats (such as prompt injection, generation of harmful content, and privacy leaks~\cite{liu2023prompt, greshake2023not, owasp2025_llm01}) require specialized detection techniques. Traditional security testing continues to play a supporting role at the infrastructure level (e.g., vulnerability scanning, access-control verification).

\paragraph{\textbf{2. Translational adaptation: semantic reconstruction of testing concepts}}
Some challenges require reinterpreting traditional testing concepts within the semantic space of LLMs. The \textit{semantic evaluation and behavioral consistency} challenge exemplifies this shift: although binary expected-versus-actual checks are inadequate, the objective of correctness validation remains. Assertions should be reformulated as multidimensional semantic evaluations that employ LLM-as-judge scoring or embedding-based semantic-similarity measures~\cite{zheng2023judging, liu2023g}.

Similarly, the \textit{capability evolution and regression risks} challenge extends regression testing to multi-task capability assessment.
Rather than detecting functional defects, we benchmark reasoning, comprehension, and generation and compare performance across model versions to identify degradation.

\paragraph{\textbf{3. Integrative innovation: deep synthesis of cross-domain methods}}
Some challenges require true methodological synthesis. For \textit{test data and robustness in an open input space}, boundary-value analysis and equivalence partitioning should be combined with adversarial-prompt generation and mutation-based perturbations~\cite{yang2024assessing, xhonneux2024efficient}. \textit{Multimodal and system-level integration} requires extending interface testing to end-to-end multimodal pipelines and coupling it with cross-modal semantic-consistency verification~\cite{abootorabi2025multimodal, han2024multimodal}. In \textit{semantic evaluation and behavioral consistency}, robustness testing benefits from integrating semantics-preserving equivalence transformations with AI-driven adversarial-perturbation techniques~\cite{weng2023adversarial, zou2023universal}.

\paragraph{\textbf{4. Runtime extension: continuous assurance beyond deployment}}
Certain categories, such as \textit{dynamic state and observability}, indicate that many critical issues (e.g., state contamination and knowledge drift) surface only in production. Static pre-deployment testing is not sufficient on its own; runtime quality assurance is needed.

In this mode, traditional monitoring (system-level metrics, anomaly detection) is combined with AI-oriented methods (semantic-quality monitoring, behavior-drift detection) to maintain continuous quality. For example, conversation-state isolation can be reinforced with semantic-contamination detection to preserve user separation and reduce subtle leakage.

\subsection{AICL: A Structured Protocol for Testable LLM Application}
\label{sec:aicl}
While free-form natural language facilitates rapid prototyping, it hinders testability:
it is ambiguous, hard to parse deterministically, and lacks machine-readable context.
This is a key reason why many existing testing techniques do not directly carry over.
Inspired by the work~\cite{xing2025when}, we therefore introduce the \emph{Agent Interaction Communication Language (AICL)}, a schema-driven, transport-independent protocol that
\textbf{(i)} enforces semantic precision via typed messages,
\textbf{(ii)} encodes observability and provenance into message metadata, and
\textbf{(iii)} guarantees replayability through canonical encodings and explicit priors for probabilistic reasoning.
AICL operationalizes the four paradigm shifts (Section~\ref{sec:shift}) by making probabilistic evaluation, dynamic exploration, runtime assurance, and method collaboration measurable and automatable.
AICL is not intended for end users. To make it practical for developers, auxiliary tooling (e.g., NL to AICL translators) is needed.

\paragraph{Protocol Overview.}
Each AICL message has a compact text form and a canonical Concise Binary Object Representation\footnote{https://en.wikipedia.org/wiki/CBOR} (CBOR) encoding:
\begin{verbatim}
[TYPE: CONTENT | METADATA]
\end{verbatim}

\paragraph{Core Message Types.}
AICL defines a small set of canonical message types that serve as building blocks for
agent interaction. Each type is motivated by the need for semantic clarity,
observability, and testability:

\begin{itemize}
  \item \texttt{HELLO} — session initialization and handshake; declares agent identity, capabilities, and version, enabling environment comparison in replay.
  \item \texttt{QUERY} — issues a request to another agent or tool, carrying input parameters and context identifiers in a structured form.
  \item \texttt{PLAN} — expresses a multi-step reasoning or execution plan, with explicit ordering and dependency markers.
  \item \texttt{FACT/FACTS} — states known information or environmental conditions, optionally annotated with confidence scores for probabilistic evaluation.
  \item \texttt{RESULT} — returns the output corresponding to a \texttt{QUERY} or \texttt{PLAN}, including data payloads and evaluation metadata.
  \item \texttt{ERROR} — standardized error reporting with error codes, context, and optional recovery hints, facilitating systematic failure analysis in testing.
  \item \texttt{MEMORY.STORE} — explicitly stores state or information, tagged with scope and context for controlled retrieval.
  \item \texttt{MEMORY.RECALL} — retrieves previously stored information, supporting reproducibility and context isolation during tests.
  \item \texttt{COORD.DELEGATE} — delegates a subtask to another agent or external tool, supporting collaborative multi-agent scenarios.
  \item \texttt{REASONING.(START|STEP|COMPLETE)} — marks stages in structured reasoning traces, enabling fine-grained capture of intermediate thoughts for later verification.
\end{itemize}

Each message is accompanied by core metadata; all fields are required unless noted. The field \texttt{of} is required on response-type messages, and \texttt{reasoning\_trace}, \texttt{cost}, \texttt{latency}, \texttt{sig}, and \texttt{cap} are optional.

\begin{itemize}
    \item \texttt{id}: unique identifier for traceability
    \item \texttt{ts}: timestamp of emission
    \item \texttt{ver}: protocol version
    \item \texttt{cid}: canonical conversation identifier for replay
    \item \texttt{ctx}: explicit context scope (ensuring isolation between tests)
    \item \texttt{model\_version}: agent model version for comparability
    \item \texttt{conf}: confidence score reported by the model
    \item \texttt{priors}: explicitly declared probabilistic assumptions
    \item \texttt{space}: domain or task space of the message
    \item \texttt{of}: correlation pointer to the upstream message \texttt{id} that this message responds to (e.g., a \textsc{RESULT} referring to its \textsc{QUERY}); typically absent on initial \textsc{QUERY}
    \item \texttt{reasoning\_trace}: structured markers (e.g., \textsc{REASONING.START|STEP|COMPLETE}) and trace payloads for capturing intermediate reasoning for later verification and analysis
    \item \texttt{cost}/\texttt{latency}: resource- and time-related annotations for performance baselining (e.g., token usage, tool/runtime latency, wall-clock duration)
    \item \texttt{sig}/\texttt{cap}: integrity and capability attestations for runtime assurance (e.g., cryptographic integrity checks and declared/attested capabilities of the agent or environment)
\end{itemize}

\noindent
Example: Query/Result Pattern.

%\begin{minted}{text}
%[QUERY: tool:weather_now{location:"Shanghai"}
 %| id:u!q1, ts:t(2025-08-15T02:00:00Z), ver:"1.2.0", ctx:[u!conv123]]

%[RESULT: {data:{temp_c:29, cond:"rain"}, schema:"tool:weather_now/1"}
% | id:u!r1, of:u!q1, conf:0.93, model_version:"gpt-5-20250801", ts:t(...)]
%\end{minted}

\begin{tcolorbox}[colback=gray!5,colframe=black,boxrule=0.5pt]
	\begin{verbatim}
		[QUERY: tool:weather_now{location:"Shanghai"}
		| id:u!q1, ts:t(2025-08-15T02:00:00Z), ver:"1.2.0", ctx:[u!conv123]]
		
		[RESULT: {data:{temp_c:29, cond:"rain"}, schema:"tool:weather_now/1"}
		| id:u!r1, of:u!q1, conf:0.93, model_version:"gpt-5-20250801", ts:t(...)]
	\end{verbatim}
\end{tcolorbox}

This example shows how a structured \texttt{QUERY} generates a corresponding \texttt{RESULT}, with explicit schema, confidence, and provenance metadata.  
Unlike natural language responses, the structured fields make it possible to evaluate correctness, latency, and reliability automatically.

\paragraph{Test-Oriented Features and Testing Benefits.}
AICL is designed explicitly for test harnesses. Its protocol-level features map directly to common challenges in LLM application testing:

\begin{itemize}
  \item Deterministic Replay and Reproducibility: Canonical CBOR encoding, together with identifiers such as \texttt{cid}, \texttt{priors}, and \texttt{model\_version}, ensures that past runs can be retrieved and replayed exactly.
  \item Context Isolation: Explicit \texttt{ctx} and \texttt{MEMORY.*} operations prevent hidden state leakage, enabling controlled experiments and differential testing.
  \item Built-in Evaluation Hooks: Structured fields such as \texttt{conf}, \texttt{reasoning\_trace}, cost/latency annotations, and typed \texttt{RESULT} payloads allow automated verdicts, schema-based validation, and performance baselining without ambiguity.
  \item Failure Diagnosis and Safety: Standardized \texttt{ERROR} codes, reasoning trace markers, and integrity checks (\texttt{sig/cap}) facilitate systematic error analysis and runtime assurance.
  \item Runtime Reusability: Production AICL logs can be ingested directly into offline regression suites, supporting continuous test–deploy–replay cycles.
\end{itemize}

\paragraph{Integration with the Collaborative Framework.}
AICL provides the structured substrate for the four collaborative modes of agent testing:
\begin{itemize}
  \item Preservational: messages act as stable, machine-verifiable interfaces, allowing classical unit and integration tests to be re-applied.
  \item Translational: semantic assertions can be applied directly to structured \texttt{RESULT} fields, avoiding the ambiguity of free-text evaluation.
  \item Integrative: tool calls, prompt templates, and retrieval events are normalized as \texttt{QUERY/RESULT} chains, enabling consistent coverage metrics across heterogeneous systems.
  \item Runtime Extension: production interactions expressed in AICL are natively reusable in regression testing, enabling continuous monitoring and runtime quality assurance.
\end{itemize}

IBM recently introduced the \textbf{Prompt Declaration Language (PDL)} \cite{ibm2024pdl}, which, like AICL, seeks to address the fragility of natural language interactions in LLM applications. PDL, as a YAML-based declarative language, emphasizes structured and maintainable prompt design, while AICL, as an interaction protocol, focuses on testing and verification through standardized message types, context isolation, and replayability. Thus, PDL operates mainly at the design layer, whereas AICL targets the execution and testing layer, making them complementary.
\section{Conclusion}
This paper is motivated by real-world quality-assurance demands of LLM applications and offers an in-depth analysis of the challenges and opportunities in this emerging field. A systematic review of the current state of AI testing reveals a structural disconnection between the software engineering and AI research communities. By analyzing fundamental differences between LLM applications and traditional software, we identify six core categories of testing challenges. By decomposing the LLM architecture into three functional layers, we discuss and analyze the applicability and evolutionary trajectory of conventional testing methods. We propose a collaborative testing framework and an agent interaction communication language~(AICL), offering a practical, systematic approach to quality assurance for LLM applications.

A key insight is that LLM application testing is neither a mere extension of traditional software testing nor a straightforward application of AI-security techniques. Rather, it is a systems-level engineering task that integrates multiple disciplinary strengths to construct a new quality-assurance paradigm. In the face of probabilistic generation, open-ended semantics, and dynamic evolution, what is required is not only technical innovation but also a shift in testing philosophy. As large language models continue to evolve and their applications proliferate, the field of LLM application testing will continue to encounter new challenges and opportunities. The analytical framework and collaborative testing architecture presented in this work provide a theoretical foundation for advancing this field, but their effectiveness should be validated and refined through empirical research, engineering practice, and cross-domain collaboration. Sustained progress will require collaboration among researchers and practitioners in software engineering, artificial intelligence, systems operations, and beyond. Advances in theory, methods, and engineering practice should proceed together to build a new quality-assurance system suited to the era of intelligent software.

\bibliographystyle{ACM-Reference-Format}
\bibliography{sample-base}

\end{document}